\documentclass[%
 reprint,
superscriptaddress,
 amsmath,amssymb,
 aps,
 nofootinbib,
]{revtex4-1}

\usepackage{dirtytalk}
\usepackage{bm}
\usepackage{amsmath,amssymb}
\usepackage{amsfonts}
\usepackage{layouts}
\usepackage{mathtools}
\usepackage{dsfont}
\usepackage{graphicx}
\usepackage{float}
\usepackage{subfigure} 
\usepackage{verbatim}
\usepackage{amsfonts}
\usepackage{bbold}
\usepackage{dcolumn}
\usepackage{bm}
\usepackage{color}
\usepackage[dvipsnames]{xcolor}
\usepackage{listings}
\definecolor{blueprl}{RGB}{46,48,146}
\usepackage[colorlinks=true,urlcolor=blueprl,citecolor=blueprl,linkcolor=blueprl]{hyperref}
\usepackage{mathrsfs}
\usepackage{filecontents}
\usepackage{times} 

\newcommand{\xdownarrow}[1]{%
  {\left\downarrow\vbox to #1{}\right.\kern-\nulldelimiterspace}
}

\makeatletter
\newcommand*{\balancecolsandclearpage}{%
  \close@column@grid
  \clearpage
  \twocolumngrid
}
\makeatother

\usepackage{gensymb}
\hypersetup{
colorlinks   = true, 
urlcolor     = blue, 
linkcolor    = blue, 
citecolor    = blue 
}
\usepackage{comment}
\usepackage[normalem]{ulem}
\usepackage{xcolor}
\usepackage{graphicx}
\usepackage{cleveref}
\crefname{equation}{Eq.}{Eqs.}
\Crefname{equation}{Equation}{Equations}
\crefname{figure}{Fig.}{Figs.}
\Crefname{figure}{Figure}{Figures}
\crefname{section}{Sec.}{Secs.}
\Crefname{section}{Section}{Sections}
\crefname{appendix}{Appendix}{Appendices}
\Crefname{appendix}{Appendix}{Appendices}
\crefname{table}{Table}{Tables}
\Crefname{table}{Table}{Tables}

\makeatletter
\def\@bibdataout@aps{%
 \immediate\write\@bibdataout{%
  @CONTROL{%
   apsrev41Control,author="08",editor="1",pages="0",title="0",year="1",eprint="1"%
  }%
 }%
 \if@filesw
  \immediate\write\@auxout{\string\citation{apsrev41Control}}%
 \fi
}%
\makeatother 

\begin{document}
\title{
Minimization of information leakage in continuous-variable quantum key distribution
}

\author{Matthew S. Winnel}\email{matthew.winnel@uqconnect.edu.au}
\affiliation{Centre for Quantum Computation and Communication Technology, School of Mathematics and Physics, University of Queensland, St Lucia, Queensland 4072, Australia}
\author{Nedasadat Hosseinidehaj} 
\affiliation{Centre for Quantum Computation and Communication Technology, School of Mathematics and Physics, University of Queensland, St Lucia, Queensland 4072, Australia}
\author{Timothy C. Ralph}
\affiliation{Centre for Quantum Computation and Communication Technology, School of Mathematics and Physics, University of Queensland, St Lucia, Queensland 4072, Australia}

\date{\today}

\begin{abstract}
A communication protocol based on a Gaussian modulation of squeezed states in a single quadrature and measured via homodyne detection can completely eliminate information leakage to an eavesdropper in a pure-loss channel. However, the asymmetry of the protocol with respect to the quadratures of light presents security issues and the eavesdropper's information is not necessarily minimized for general asymmetric attacks. Here, we perform asymptotic security analysis of the asymmetric protocol against general asymmetric collective attacks and bound the eavesdropper's information via the Heisenberg uncertainty principle. The bound is not tight and therefore, we symmetrize the protocol in a heralding way, discarding the issues of asymmetry altogether. Our proposed heralding protocol asymptotically eliminates information leakage in a pure-loss channel and minimizes leakage in a noisy channel.
\end{abstract}

\maketitle

\section{\label{sec:intro}Introduction}

Quantum key distribution (QKD)~\cite{scarani2009security,Pirandola_2020} is the task of distributing a secret random key between two distant parties. Its provable security relies on quantum physics and it is therefore a promising solution to the vulnerability of current classical cryptosystems. The first QKD protocols were based on discrete-variables (DV), such as BB84~\cite{BB84}, where the fundamental component is a qubit. Recently, QKD has been extended to continuous-variable (CV)~\cite{cerf2007quantum, weedbrook2012gaussian} systems, utilizing the infinite-dimensional space of modes of light, with the benefit of using simpler experimental set-ups and off-the-shelf optical communication devices such as homodyne detectors.



There is a zoo of CV QKD protocols~\cite{weedbrook2004quantum,gehring2015single, usenko2018unidimensional,usenko2019generalized,Grosshans_2002,cerf2001quantum,gottesman2003secure}. For instance, either coherent states~\cite{Grosshans_2002,weedbrook2004quantum} or squeezed states~\cite{cerf2001quantum,gottesman2003secure} may be distributed between the trusted parties, and the ensemble may be symmetric or asymmetric with respect to the quadratures of light. See Ref.~\cite{Pirandola_2020} for a review. A key difference between CV QKD and DV QKD is that in DV QKD lost photons are discarded, but in CV QKD all states are retained but are noisier.

A CV analogue of BB84 is the protocol based on Gaussian modulated squeezed states with switching (i.e., randomly
choosing to squeeze and modulate either the $x$ or $p$ quadrature) and measured via homodyne detection~\cite{cerf2001quantum}, however, even in a pure-loss channel, information is inevitably leaked to an eavesdropper. During the protocol, correlations exist between all three parties, and the trusted parties, Alice and Bob, must suppress their correlations with the eavesdropper, known as Eve, to ensure the final key is secret. This can be done via error correction, privacy amplification, or for an entanglement protocol, via entanglement distillation or purification. Recently, a CV QKD protocol has been devised which takes a different approach. By designing the alphabet of input states in a certain way, information leakage to Eve is completely and deterministically eliminated in a pure-loss channel and minimized in a symmetric noisy channel~\cite{jacobsen2018complete}. Information is encoded in a single quadrature via a Gaussian modulation of squeezed states, which are squeezed in the modulation direction such that the overall variance of the ensemble is shot noise. The motivation for minimizing information leakage is that this will simplify classical post-processing and ultimately speed up the secret key rate. Another motivation is that zero information leakage in a pure-loss channel is analogous to discarded photons in DV QKD.


It is vital that QKD protocols can be proved secure against the most general eavesdropping attacks allowed by the laws of quantum physics. Indeed, currently composable security for finite key lengths has been proved against general attacks for only several CV QKD protocols, such as the no-switching protocol based on Gaussian modulated coherent states and heterodyne detection~\cite{leverrier2017security}, and the Gaussian modulated squeezed state protocol with switching and homodyne detection~\cite{furrer2011continuous}. Recently, a discrete modulation protocol has been proved secure including finite-size effects~\cite{matsuura2020finitesize}. In Ref.~\cite{jacobsen2018complete}, the authors considered a symmetric Gaussian attack which is not asymptotically optimal since the protocol is asymmetric and the channel parameters in the unmodulated quadrature are unknown.

In this paper we deal with the asymmetry of the asymmetric minimum-leakage protocol from Ref.~\cite{jacobsen2018complete}. We first extend their security analysis to include general asymmetric Gaussian attacks. We show that Eve's information is not necessarily minimized for asymmetric attacks. Further, we introduce a new protocol which is strictly symmetric with respect to the quadratures of light. This forces Eve to implement symmetric attacks thereby avoiding the issues of asymmetry. Conditioned on a homodyne measurement at Alice's station, our protocol heralds squeezed states modulated in a single quadrature. We show that for this new protocol information leakage is asymptotically eliminated for pure-loss channels, and minimized for noisy channels. 


%

%



The outline of this paper is as follows. In \cref{sec:security}, we revise state of the art security for CV QKD, and we introduce the asymptotic key rate formula and the Holevo bound. In \cref{sec:asymmetric}, we recall the asymmetric protocol from Ref.~\cite{jacobsen2018complete} and extend their analysis to include general asymmetric attacks. In \cref{sec:herald}, we introduce a symmetric heralding protocol. In \cref{sec:discussion}, we discuss our results and conclude.

\section{Security of CV QKD protocols\label{sec:security}}

\subsection{\label{sec:challenges}State-of-the-art security proofs}

The goal of security analysis is to provide a useful upper bound on Eve's information using correlations between Alice and Bob, leading to an expression for the secret key rate. Unfortunately, proving security is problematic in CV QKD because the Hilbert space is infinite dimensional, the measurement operators are unbounded, and channel parameters must be estimated of noisy quantum channels. Compare this with DV QKD where the Hilbert space is finite dimensional and loss can be dealt with by simply discarding the state. It is known that Gaussian collective attacks are asymptotically optimal for fully Gaussian CV QKD protocols~\cite{Garcia-Patron2006unconditional}, i.e. in the asymptotic case of infinitely long keys. To go beyond asymptotic analysis, only a few Gaussian CV QKD protocols are known to be composably secure against general attacks and including finite-size effects, for instance, the no-switching protocol based on Gaussian modulated coherent states in the $x$ and $p$ quadratures and heterodyne detection~\cite{leverrier2017security,Ghorai_2019}, and the switching protocol based on squeezed states with a Gaussian modulation in the $x$ or $p$ quadratures and homodyne detection such that the overall state prepared by Alice is a Gaussian thermal state ~\cite{furrer2011continuous,furrer2014reverse}.


We restrict our attention here to the asymptotic regime though we caution that finite-size effects are expected to be significant. The original asymmetric protocol from Ref.~\cite{jacobsen2018complete} with squeezed states is analyzed against collective attacks in the asymptotic limit in~\cref{sec:asymmetric}. Our symmetric heralding protocol, discussed in~\cref{sec:herald}, is Gaussian and symmetric, however, current composable security proofs including finite-size effects still fail due to Alice's homodyne detection. 
%

\subsection{\label{sec:CVQKD}Asymptotic secret key rate formula and Eve's maximal information}

A lower bound on the asymptotic secret key rate in the case of reverse reconciliation for collective attacks is given by the Devetak-Winter rate~\cite{devetak2005distillation}
\begin{equation}
K = \beta I_{AB} - \chi_{EB},
\end{equation}
where $I_{AB}$ is the classical (Shannon) mutual information between Alice and Bob, $\chi_{EB}$ is the Holevo quantity, the maximal quantum mutual information between Eve and Bob (the reference side of the information reconciliation), and $\beta$ is the reconciliation efficiency.

The upper bound on the information extractable by Eve is given by the Holevo quantity~\cite{holevo1973bounds}
\begin{equation}
\chi_{EB} = S(\rho_E) - S(\rho_{E|b}),\label{eq:HOLEVO}
\end{equation}
where $S(\rho_E)$ is the von Neumann entropy of Eve's state, and $S(\rho_{E|b})$ is the von Neumann entropy of Eve's state conditioned on Bob's measurement. 

The Holevo information is obtained by allowing Eve access to the purification of the state shared between Alice and Bob $\rho_{AB}$. The global state $\rho_{ABE}$ is pure and we can use the self-duality property of the von Neumann entropy to write $S(\rho_E)=S(\rho_{AB})$ and $S(\rho_{E|b})=S(\rho_{A|b})$, where $S(\rho_{A|b})$ is Alice's mode conditioned on a measurement on Bob's mode. 

For Gaussian states, $\chi_{EB}$ may be calculated from the symplectic eigenvalues of the covariance matrices of $\rho_{AB}$ and $\rho_{A|b}$ in the equivalent entanglement-based (EB) version of a prepare-and-measure CV QKD protocol. Thus, Eve's information is given by
\begin{equation}
\chi_{EB} = S(\rho_{AB}) - S(\rho_{A|b}),
\end{equation}
i.e. the Holevo quantity is given in terms of von Neumann entropies which can be calculated using the symplectic eigenvalues $v_i$ of the covariance matrix of the state, i.e. $\Gamma_{AB}$ and $\Gamma_{A|b}$, via the relation $S(\rho)=\sum_{i=1}^N \frac{v_i+1}{2}\log_2 \frac{v_i+1}{2} - \frac{v_i-1}{2}\log_2 \frac{v_i-1}{2},$ where $N$ is the number of modes.

\section{\label{sec:asymmetric}Asymmetric protocol}

\begin{figure}
    \hfill
    \centering
    \begin{subfigure}[]
        \centering
        \includegraphics[width=1\linewidth]{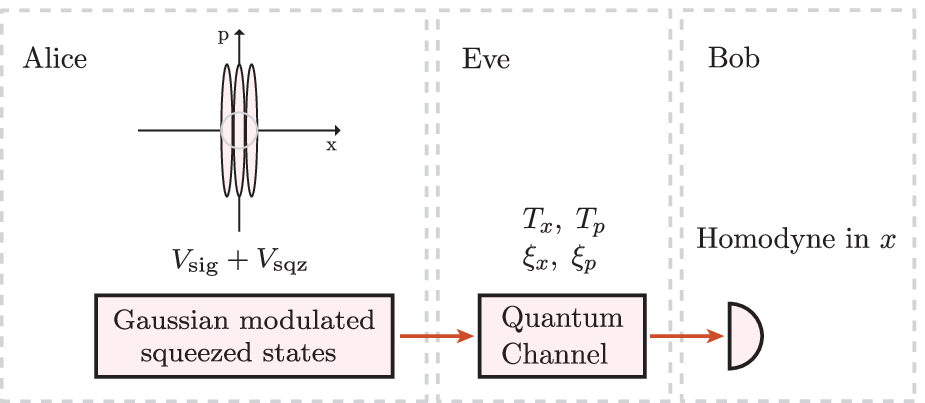}
    \end{subfigure}
        \begin{subfigure}[]
        \centering
        \includegraphics[width=1\linewidth]{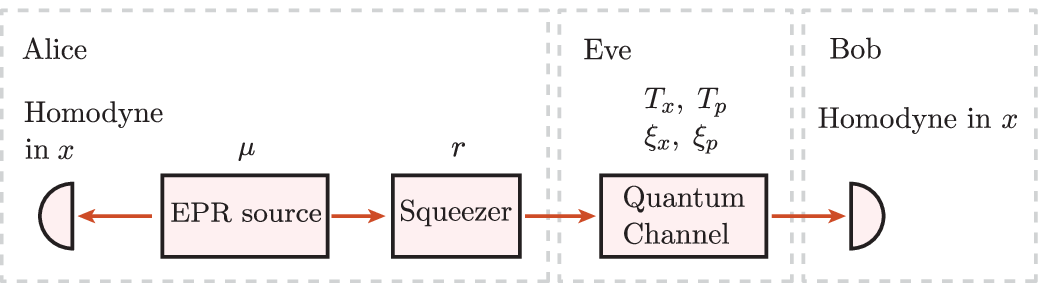}
    \end{subfigure}
\caption{Equivalent schemes of the asymmetric minimum-leakage protocol from Ref.~\cite{jacobsen2018complete} a) prepare-and-measure version and b) entanglement-based version. The asymmetric protocol consists of squeezed states modulated and squeezed in the squeezing direction. When the variance of the overall ensemble is shot noise, information leakage is zero in a pure-loss channel.}\label{fig:QKDschemes}
\end{figure}

The asymmetric minimum-leakage protocol from Ref.~\cite{jacobsen2018complete} is shown in~\cref{fig:QKDschemes}. It is based on squeezed states modulated in a single direction with an overall Gaussian modulation and squeezing chosen so that the Holevo information $\chi_{EB}$ is eliminated or minimized while the key rate is non-zero. Let us assume that the modulated quadrature is chosen to be the amplitude quadrature $x$. In a pure-loss channel, the requirement for complete elimination of $\chi_{EB}$ is for the ensemble of squeezed states to have an overall amplitude quadrature noise variance of vacuum.

In the prepare-and-measure (PM) version of the protocol (the version Alice and Bob implement in practice), Alice prepares a Gaussian modulation of squeezed states of light to send to Bob. The PM scheme of the protocol is given in \cref{fig:QKDschemes}(a). Alice prepares $x$ squeezed states with amplitude quadrature variance $V_{\text{sqz}}$ and applies modulation in the $x$ quadrature, displacing each squeezed state according to a random Gaussian variable with variance $V_{\text{sig}}$. The states are then sent to Bob through an asymmetric noisy channel with transmittance $T_x,\;T_p$ and excess noise $\xi_x,\;\xi_p$ in the $x$  and $p$ quadratures respectively. Bob performs homodyne measurements of the modulated $x$ quadrature, but sometimes also measures the unmodulated $p$ quadrature for estimating the properties of the channel in the $p$ quadrature. Alice and Bob extract a secret key from the $x$ quadrature data using a reverse-reconciliation procedure.

Security is analyzed using the equivalent EB version of the protocol, shown in~\cref{fig:QKDschemes}b) which goes as follows. Alice prepares a two-mode squeezed vacuum state of variance $\mu$, keeps one of the modes, and squeezes the $x$ quadrature of the second mode with squeezing parameter $r$ before sending it to Bob. Alice performs homodyne measurements in the $x$ quadrature in order to project Bob's mode onto an ensemble of squeezed states with a Gaussian modulation. For instance, if Alice homodynes $x$ then effectively she has sent a Gaussian modulation of $x$ squeezed states in the $x$ direction to Bob, equivalent to the PM version.

As shown in Ref.~\cite{jacobsen2018complete}, the PM condition which completely decouples the eavesdropper in a pure-loss channel is $V_\text{sig}+V_\text{sqz} = 1$, an overall $x$ quadrature noise variance of vacuum. One can arrive at this condition by calculating Eve's maximal information via the covariance matrix shared between Alice and Bob in the EB version, and we refer you to \cref{sec:appendix:symmetricattack} and Ref.~\cite{jacobsen2018complete} for details. A similar relation minimizes the Holevo information under the assumption of a symmetric noisy channel but a little more squeezing is required because of the noise. By symmetric channel we mean $T_x = T_p$ and $\xi_x = \xi_p$, which is a restricted eavesdropping attack since the input state to the channel is asymmetric. The dashed curve in~\cref{fig:uniprotocol} shows that Eve's information is minimized in a symmetric noisy channel for an appropriate choice of $V_\text{sqz}$ given that $V_\text{sig}=0.5$. In the next section we go beyond symmetric attacks.

\subsection{General asymmetric attacks}

In this section, we consider channels where the noise and loss may be asymmetric in each of the quadratures. Recall that during the protocol, Alice and Bob perform parameter estimation to derive the covariance matrix shared between them. Usually both quadratures are modulated allowing the full covariance matrix to be estimated, however, although only one quadrature is modulated, Bob can still estimate the variance of the unmodulated quadrature and use physicality of the quantum state to bound Eve's information, i.e., via the Heisenberg uncertainty principle~\cite{usenko2018unidimensional}.


\subsubsection{\label{sec:Heisenberg}Bounding Eve using the Heisenberg Uncertainty Principle}


We use the EB version shown in \cref{fig:QKDschemes} b) and bound Eve using the Heisenberg uncertainty principle. The EB parameters are related to the PM parameters such that
\begin{align}
\begin{split}
    \mu V&=V_\text{sig}+V_\text{sqz}\\
    \frac{\mu}{V}&=\frac{1}{V_\text{sqz}},\label{eq:EBparamaters1}
\end{split}
\end{align}
where $\mu$ is the EPR variance and $V=e^{-2r}$ is the strength of the squeezing on the outgoing mode with squeezing parameter $r$. Equivalently, solving for $\mu$ and $r$, we have
\begin{align}
\begin{split}
    \mu&=\sqrt{1+\frac{V_\text{sig}}{V_\text{sqz}}}\\
    r&=-\frac{1}{2}\ln(\sqrt{V_\text{sqz}(V_\text{sqz}+V_\text{sig})}).\label{eq:EBparamaters2}
\end{split}
\end{align}
The covariance matrix shared between Alice and Bob after Eve's asymmetric attack is given by (see \cref{sec:appendix:asymmetricattack} for details):
\begin{align}
\Gamma_{AB} &= \left[ \begin{smallmatrix} \mu & 0 & e^{-r}\sqrt{T_x (\mu^2-1)} & 0\\ 0 & \mu & 0 & c_p \\ e^{-r}\sqrt{T_x (\mu^2-1)}  & 0 & T_x (e^{-2r}\mu+\xi_x)+1-T_x & 0 \\ 0 & c_p & 0 &v_p^B\end{smallmatrix} \right],
\end{align}
where $v_p^B$ and $c_p$ are unknown since the $p$ quadrature is unmodulated. Fixing $T_x$ and $\xi_x$, then given Bob's $p$ quadrature variance  $v_p^B$ measured during an experiment (which he only does sometimes since he is mostly performing a homodyne measurement of the $x$ quadrature), we bound the unknown correlation parameter $c_p$ by the physical requirement of the state given by $\Gamma_{AB} + i\Omega \geq 0$, where $\Omega$ is the symplectic form $\Omega = \bigoplus_{i=1}^{n} \omega,\;\omega = \left( \begin{smallmatrix} 0 & 1 \\ -1 & 0 \end{smallmatrix} \right)$.

To calculate a general upper bound on Eve's information we simulate an experiment and give Bob's variance $v_p^B$ the value it would have if the channel were symmetric (i.e., in our simulated experiment $v_p^B$ is calculated using $T_p=T_x$ and $\xi_p=\xi_x$), and then we maximize Eve's information $\chi_{EB}$ by going over all physical covariance matrices.  This is plotted in \cref{fig:uniprotocol} (solid). Also plotted is for a symmetric channel (dashed), which has been done in~\cite{jacobsen2018complete}. Note that Alice prepares a Gaussian modulation of coherent states for no squeezing $V_\text{sqz}=1$. For coherent states, bounding Eve's information in this way is not very pessimistic, as can be seen in the plot since the gap between symmetric and upper bound is not significant. But as squeezing is increased, the gap between the symmetric channel and the general upper bound becomes very significant. To provide tighter bounds, it is better to measure all terms of the covariance matrix, thus, we need to do some estimation of the unmodulated quadrature.

\subsubsection{\label{sec:estimatep}Estimating the unmodulated quadrature}

\begin{figure}
    \hfill
        \centering
        \includegraphics[width=1\linewidth]{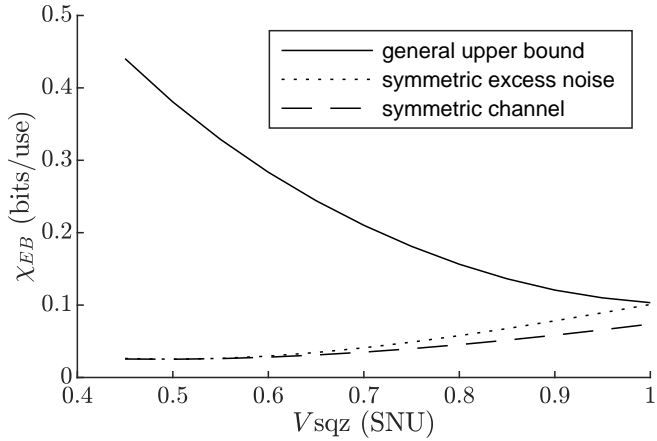}
    \caption{Bound on Eve's information versus squeezing for the asymmetric protocol with transmittance $T_x=0.5$, excess noise $\xi_x=0.01$ SNU, and $V_\text{sig}=0.5$. The upper bound becomes very loose with squeezing. Shown are symmetric lossy thermal channel (dashed), general upper bound (solid), and equal excess noise in both quadratures (dotted).}\label{fig:uniprotocol}
\end{figure}

One must perform some estimation of the unmodulated quadrature in an experiment to obtain the full covariance matrix but it is difficult to estimate the unmodulated quadrature since it is antisqueezed. For illustration purposes, we assume that some estimation of the $p$ quadrature is performed and that the excess noise is equal in both quadratures, $\xi_p = \xi_x$, and we again bound Eve's information using physicality of the covariance matrix and plot this in \cref{fig:uniprotocol} (dotted). We do not have to simulate Bob's variance in this case because there is only one unknown parameter, $T_p$. Compared to the general upper bound, assuming equal excess noise in both quadratures does much better. This shows that it is important to estimate the noise in the unmodulated quadrature, however, equal excess noise in both quadratures is not a realistic assumption because in an experiment the noise associated with the unmodulated quadrature is expected to be worse than the modulated quadrature. Therefore, one can expect Eve's information to lie between the symmetric channel and the upper bound. In summary, Eve's information is not necessarily minimized when considering stronger attacks beyond symmetric ones (and note that we have not yet considered finite-size effects). This motivates an exploration of symmetric protocols with the goal of minimizing Eve's information.


\section{\label{sec:herald}Heralding protocol}

\begin{figure}
    \hfill
    \centering
    \begin{subfigure}[]
        \centering
        \includegraphics[width=1\linewidth]{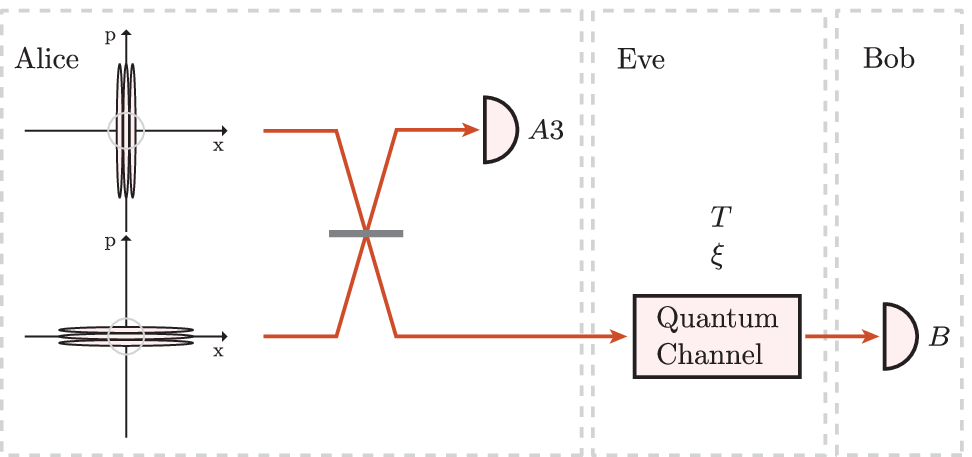}
    \end{subfigure}
        \begin{subfigure}[]
        \centering
        \includegraphics[width=1\linewidth]{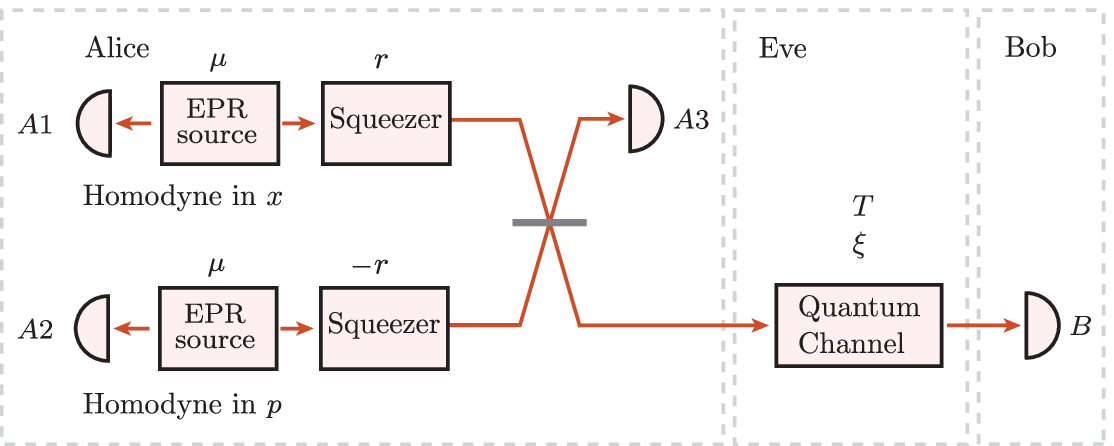}
    \end{subfigure}
\caption{Equivalent schemes of the heralding protocol a) prepare-and-measure version and b) entanglement-based version. Alice and Bob randomly and independently homodyne either $x$ or $p$ at both modes $A3$ and $B$ and later sift the results. The protocol consists of all Gaussian elements and is symmetric with respect to the quadratures, and eliminates information leakage in a pure-loss channel when~\cref{eq:heralding} is satisfied. See~\cite{Madsen2012} for a proof of principle experiment.}\label{fig:QKDschemesheralding}
\end{figure}

In this section,  we symmetrize the asymmetric protocol of Ref.~\cite{jacobsen2018complete} in a heralding way to avoid worrying about asymmetric attacks in the security analysis. We present our heralding protocol in~\cref{fig:QKDschemesheralding}. In the PM version, Alice prepares two modes; two independent ensembles of squeezed states with a Gaussian modulation in a single direction, one mode modulated in the $x$ direction and the other mode in the $p$ direction. Note that here again we use $V_\text{sqz}$ for the squeezing variance and $V_\text{sig}$ for the modulation variance. These are combined on a beamsplitter and Alice homodynes one of the outputs (labelled $A3$) in either $x$ or $p$ at random while the other mode is sent to Bob (labelled $B$). The protocol heralds squeezed states modulated in a single quadrature, conditioned on Alice's homodyne measurement at $A3$. Bob independently and randomly homodynes in either $x$ or $p$ and, during classical post-processing, Alice and Bob sift their results, where they only keep the data for which they have used the same quadrature for the measurement.

In the EB version, Alice's ensembles are replaced with EPR states plus auxiliary squeezing on one of the modes. Mode $B$ is squeezed in the $x$ quadrature with squeezing parameter $r$ and mode $A3$ is squeezed in the $p$ direction by an equal amount. The EB version is equivalent to the PM version if Alice homodynes $A1$ in $x$ and $A2$ in $p$.


Our heralding protocol consists of all Gaussian elements and is symmetric, so we can analyze security against Gaussian collective attacks in the asymptotic limit. For our heralding protocol we do know that in the asymptotic limit Gaussian attacks are optimal, and, unlike the asymmetric protocol, we no longer have to worry about asymmetric attacks since the protocol is symmetric. We still have the issue that we do not know Eve's optimal attack in the finite-size regime.

Now we consider the EB version of the Heralding protocol. Before going through the channel, the mode that will go to Bob is symmetric and is given by
\begin{align}
\Gamma_B &= \left[  \begin{smallmatrix} \mu  \cosh{2r} &0 \\ 0 & \mu \cosh{2r}  \end{smallmatrix}  \right].
\end{align}

After the channel, homodyne detection is performed on modes $A3$ and $B$ independently and randomly in $x$ and $p$. At the end Alice and Bob sift, and here, without loss of generality, we assume this sifting is in the $x$ quadrature (shifted results in the $p$ quadrature will behave similarly). After the quantum channel and Alice's heralding measurement (which we assume is homodyne $A3$ in $x$), the covariance matrix of Bob's mode is
\begin{align}
\Gamma_{B|A3_x} &= \left[  \begin{smallmatrix}   \frac{e^{4r} - T + T\xi - Te^{4r} + T\xi e^{4r} + 2T\mu e^{2r} + 1}{e^{4r} + 1} &0 \\ 0 & T\xi - T + T\mu \cosh{2r} + 1  \end{smallmatrix} \right].
\end{align}
The full covariance matrix is given in~\cref{sec:appendix:heralding}.

For zero information leakage in a pure-loss channel we require that the variance of the $x$ quadrature is equal to the shot noise. Solving for the prepare-and-measure parameters defined in~\cref{eq:EBparamaters1}, we find the following condition
\begin{align}
V_\text{sig} &= \frac{{V_\text{sqz}}^2 -2 V_\text{sqz} +1}{2 - V_\text{sqz}}.
\label{eq:heralding}
\end{align}

\begin{figure}
    \centering
    \begin{subfigure}[]
        \centering
        \includegraphics[width=1\linewidth]{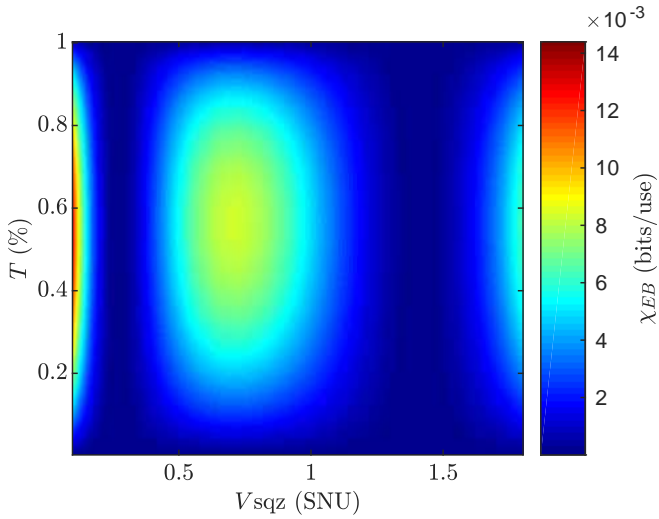}
    \end{subfigure}
    \begin{subfigure}[]
        \centering
        \includegraphics[width=1\linewidth]{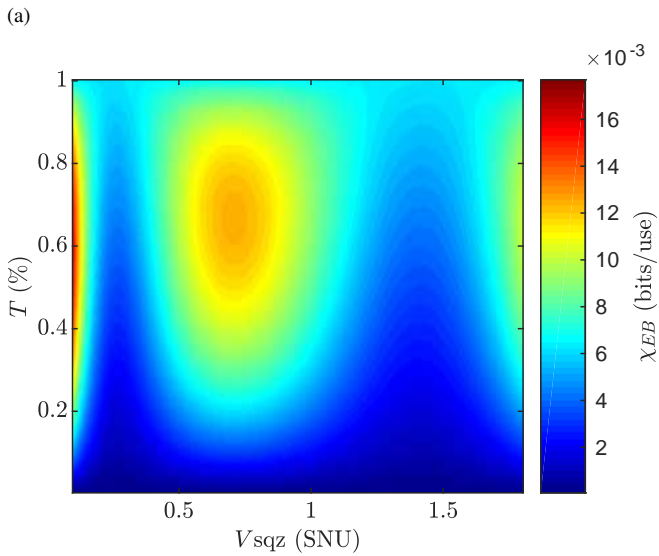}
    \end{subfigure}
\caption{Eve's information for the heralding protocol as a function of squeezing variance $V_\text{sqz}$ and transmissivity $T$ with fixed modulation variance $V_\text{sig}=0.3$: a) no excess noise, b) excess noise $\xi = 0.001$ SNU. In a pure-loss channel Eve's information can be zero, and with added thermal noise Eve's information is minimized. The region of interest is when $V_\text{sqz}<1$.} 
\label{fig:heraldingcontour}
\end{figure}

For example, if $V_\text{sig}=0.3$, then this condition implies that there are two solutions which eliminate the Holevo quantity: squeezing $V_\text{sqz} = 0.2821$ or antisqueezing $V_\text{sqz}=1.4179$. We are interested in the squeezing solution, not antisqueezing (since mutual information $I_{AB}$ decreases with antisqueezing), and note that $V_\text{sqz}=0$ for infinite squeezing and $V_\text{sqz}=1$ for zero squeezing. \Cref{fig:heraldingcontour} shows Eve's information as a function of transmissivity $T$ and squeezing variance $V_\text{sqz}$ for a) pure-loss and b) excess-noise $\xi = 0.001$. Eve's information is eliminated in a pure-loss channel when \cref{eq:heralding} is satisfied. For the noisy case~\cref{fig:heraldingcontour} b) shows that when~\cref{eq:heralding} is approximately satisfied and the Holevo information is minimized, the Holevo information actually decreases with loss. In other words, the Holevo information goes to zero as the transmissivity of the channel is decreases. That Eve's information is less the more loss there is may be counter-intuitive (since she steals more but gets less) but makes sense when one remembers that Alice and Bob's mutual information is also less for more loss.  

\begin{figure}
    \centering
        \includegraphics[width=1\linewidth]{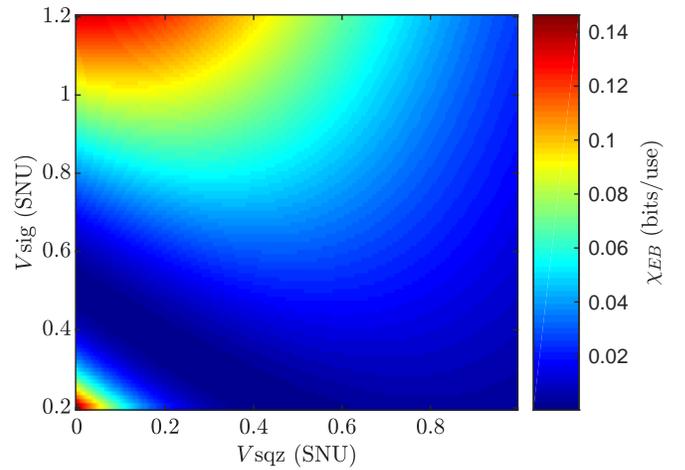}
\caption{Eve's information for the heralding protocol as a function of squeezing variance $V_\text{sqz}$ and signal variance $V_\text{sig}$ in a pure-loss channel with fixed transmissivity $T=0.5$. As $V_\text{sqz}\to0$ Eve's information is zero when $V_\text{sig}\to0.5$. Coherent states are prepared at Alice when $V_\text{sqz}{=}1$ and Eve's information cannot be zero in this case for any choice of modulation.}
\label{fig:heraldingcontour2}
\end{figure}

In~\cref{fig:heraldingcontour2}, we plot Eve's information as a function of signal modulation and squeezing in a pure-loss channel with fixed transmissivity $T=0.5$. This shows that as squeezing is increased from $V_\text{sqz}=1$ (coherent states) to $V_\text{sqz}\to0$ (infinite squeezing), the modulation variance $V_\text{sig}$ must also be increased in order to minimize information leakage.

\begin{figure}
    \centering
        \centering
        \includegraphics[width=1\linewidth]{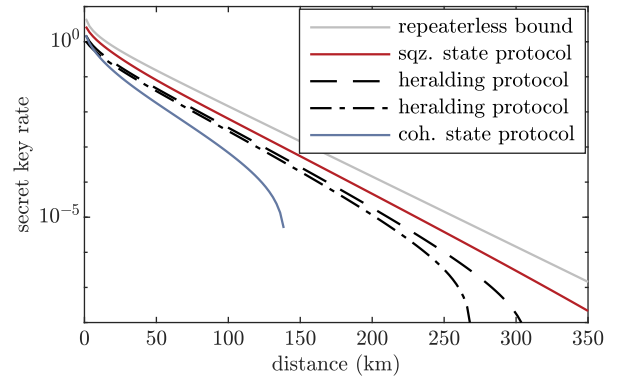}
\caption{Secret key rate of the heralding protocol as a function of distance for lossy thermal channel. The dashed and dot-dashed lines (black) show the heralding protocol with infinite and 10dB of finite squeezing respectively. The heralding protocol optimizes at very large squeezing, i.e., $V_\text{sqz}{\to}0,\;V_\text{sig}{\to}0.5$. For these plotted rates, the condition \cref{eq:heralding} is satisfied at all distances, hence, Eve's information is zero at all distances in a pure-loss channel and small in a noisy channel. Shown for comparison are optimized rates for the original squeezed state protocol with switching and homodyne detection~\cite{cerf2001quantum} and no-switching coherent state protocol with heterodyne detection~\cite{weedbrook2004quantum}. The excess noise is $\xi = 0.05$ for all protocols, the reconciliation efficiency is $\beta{=}0.95$ and we have considered optical fiber with loss of 0.2dB/km. The repeaterless PLOB bound \cite{Pirandola_2017} and is also plotted in the figure (gray).}
\label{fig:heraldingrates}
\end{figure}

In~\cref{fig:heraldingrates}, we show the secret key rate of the heralding protocol as a function of distance for a thermal-loss channel with excess noise $\xi = 0.05$. Alice and Bob's mutual information is calculated assuming modes $A1$, $A3$ and $B$ are all homodyned in $x$, and the parameters of the protocol are fixed such that \cref{eq:heralding} is satisfied and Eve's information is eliminated or approximately minimized at all distances. Alice and Bob's mutual information is calculated after Alice's heralding measurement of mode $A3$ in the $x$ quadrature and is given by
\begin{align}
I_{AB} = \frac{1}{2} \log_2 \frac{V_{B_x}}{V_{B_x|A1_{\text{hom}x}}},
\end{align}
where $V_{B_x}$ is the variance of Bob's $x$ quadrature, since this is the quadrature we assume he uses for the key, and ${V_{B_x|A1_{\text{hom}x}}}$ is the variance of Bob's $x$ quadrature conditioned on Alice's homodyne measurement in $x$ of mode $A1$.

The heralding protocol optimizes for very large squeezing shown as the dashed line in~\cref{fig:heraldingrates}. Infinite squeezing is $V_\text{sqz}{\to}0$ and the zero-leakage condition~\cref{eq:heralding} means that $V_\text{sig}{\to}0.5$. Also shown is the key rate for the heralding protocol with 10dB of finite squeezing (dot-dashed). For comparison, we plot the optimized key rate for Gaussian-modulated squeezed state protocol and measured via homodyne detection~\cite{cerf2001quantum} (red) and Gaussian-modulated coherent state protocol and heterodyne detection~\cite{weedbrook2004quantum} (blue). The optimized squeezed state protocol with homodyne detection outperforms our heralding protocol. However, since our heralding protocol likewise is based on squeezed states and homodyne detection it can outperform the coherent state protocol with heterodyne detection.

We note here that rates for the heralding protocol are numerically identical as those for the asymmetric protocol against symmetric thermal-loss channel. Indeed, the rate for the heralding protocol for a pure-loss channel, infinite squeezing, and reconciliation efficiency $\beta{=}1$ reaches half of the fundamental repeaterless PLOB bound, which is given by $K_\text{PLOB} = {-}\log_2{(1-\eta)}$ \cite{Pirandola_2017}. This result for the asymmetric protocol was indeed mentioned in~\cite{jacobsen2018complete}.


\section{Discussion and conclusion\label{sec:discussion}}

In this paper, we have investigated CV QKD protocols designed to eliminate information leakage to Eve. We extended security analysis of the original asymmetric minimum-leakage protocol by considering general asymmetric channels by demanding physicality of the quantum states. We also introduced a new protocol by symmetrizing in a heralding way.

The main results from this paper are the following: if the unmodulated quadrature can be properly estimated, the asymmetric protocol minimizes Eve's information and is secure against Gaussian attacks in the asymptotic regime. However, using the Heisenberg uncertainty principle for security analysis can give Eve a lot of information. Our heralding protocol minimizes Eve's information in a heralding way despite being symmetric with respect to the quadratures, but with added experimental complexity.

Another way to symmetrize the asymmetric minimum-leakage protocol is by switching, i.e., randomly sending $x$-squeezed or $p$-squeezed states with an overall modulation of shot noise in each quadrature. However, the overall ensemble prepared by Alice will be non-Gaussian (specifically, it will not be a Gaussian thermal state as in the squeezed state protocol with homodyne detection from Ref.~\cite{cerf2001quantum}). The non-Gaussianity complicates the security analysis, potentially giving more information to the eavesdropper, departing from our goal of minimum-leakage. We leave the investigation of minimum-leakage switching protocols for future work.

In conclusion, we have introduced a new CV QKD protocol which minimizes the amount of information leaked to an eavesdropper in a noisy channel and eliminates information leakage in a pure-loss channel, meaning that the classical part of the protocol will be computationally less complex, potentially leading to an overall increase in the secret key rate in situations of practical interest.



\clearpage

\begin{acknowledgements}
We thank Iyad Suleiman, Tobias Gehring, Jonas S. Neergaard-Nielsen, and Ulrik L. Andersen for valuable discussions during the investigation of this work. This research was supported by the Australian Research Council (ARC) under the Centre of Excellence for Quantum Computation and Communication Technology.
\end{acknowledgements}

\onecolumngrid
\appendix

\section{\label{sec:appendix:asymmetric}Asymmetric protocol}

In this section we write down the covariance matrix of the asymmetric minimum-leakage protocol from Ref.~\cite{jacobsen2018complete} in the EB version. The goal is to calculate the Holevo bound after symmetric and asymmetric Gaussian noisy channels. The covariance matrix before the channel shared between Alice and Bob is
\begin{align*}
\Gamma_{\text{input}} &= \left[ \begin{smallmatrix} 
                       \mu &                        0 & e^{-r}\sqrt{\mu^2 - 1} &                        0 \\
                        0 &                       \mu &                        0 & -e^{r}\sqrt{\mu^2 - 1} \\
 e^{-r}\sqrt{\mu^2 - 1} &                        0 &             \mu e^{-2r} &                        0 \\
                        0 & -e^{r}\sqrt{\mu^2 - 1} &                        0 &              \mu e^{2r} \\
\end{smallmatrix} \right],
\end{align*}
where $\mu$ is the strength of the EPR state and $r$ is the amount of the squeezing.

If the EB squeezing parameter is $r=-\text{ln}\sqrt{\mu}$ (the minus sign means squeezing in the $p$ direction), then Alice has effectively sent coherent states to Bob in the PM version. For other amounts of squeezing, Alice has conditionally prepared squeezed states in the PM version. Specifically, for the squeezing parameter $r=\text{ln}\sqrt{\mu}$ (squeezing in the $x$ direction), the modulation variance in the $x$ quadrature is shot noise and this is the PM condition $V_\text{sig}+V_\text{sqz}=1$ which eliminates Eve's information for a pure-loss channel. If we choose to modulate in the $x$ quadrature, then the covariance matrix in the EB version for zero Holevo in a pure-loss channel is
\begin{align}\label{eq:shotnoiseCM}
\Gamma_{AB} &= \left[ \begin{smallmatrix} \mu & 0 & \sqrt{\frac{\mu^2-1}{\mu}} & 0\\ 0 & \mu & 0 & -\sqrt{\mu(\mu^2-1)} \\ \sqrt{\frac{\mu^2-1}{\mu}}   & 0 & 1 & 0 \\ 0 & -\sqrt{\mu(\mu^2-1)} & 0 & \mu^2\end{smallmatrix} \right].
\end{align}
There is a symmetry between the zero information leakage squeezed state protocol and the coherent state single quadrature protocols. It arises because of the requirement to squeeze to the shot noise, and coherent states have an intensity noise at the shot noise level by definition. Given the covariance matrix in \cref{eq:shotnoiseCM}, if Alice chooses to homodyne in $x$ then $x$ modulated squeezed states are conditionally prepared at Bob. If Alice homodynes in $p$, $p$ modulated coherent states are prepared at Bob. The Holevo quantity is eliminated only if Bob chooses to homodyne in the quadrature at the shot noise, i.e. Alice and Bob homodyne the same quadrature. 

Midway between the two EB protocols (i.e. midway between coherent state and shot noise $r=\pm{}\text{ln}\sqrt{\mu}$), Alice performs no squeezing on Bob's mode in the EB version $r=0$, then the EB protocol is equivalent to the original squeezed state protocol with switching which is symmetric between the two quadratures \cite{cerf2001quantum}.

\subsection{\label{sec:appendix:symmetricattack}Symmetric attack}

Let's assume the channel is symmetric. After the symmetric channel the covariance matrix shared between Alice and Bob is
\begin{align*}
\Gamma_{AB} &= \left[ \begin{smallmatrix}  
                               \mu &                                0 & c_x &                                0 \\
                                0 &                               \mu &                                0 & c_p \\
c_x &                                0 &     v_x^B &                                0 \\
                                0 & c_p &                                0 &  v_p^B \\
 \end{smallmatrix} \right],
 \end{align*}
 with 
 \begin{align*}
 c_x = e^{-r}\sqrt{T(\mu^2 - 1)}\\
  c_p = -e^{r}\sqrt{T(\mu^2 - 1)}\\
 v_x^B = T(e^{-2r}\mu +\xi) + 1 - T\\
v_p^B = T(e^{2r}\mu +\xi) + 1 - T,
 \end{align*}
where $T$ is the transmissivity and $\xi$ is excess noise, symmetric in both the $x$ and $p$ quadratures.

The covariance matrix for Alice's mode conditioned on Bob's homodyne measurement, which is assumed to be in the $x$ quadrature, is
 \begin{align*}
\Gamma_{A|b} &= \left[ \begin{smallmatrix} 
 \mu - \frac{Te^{-2r}(\mu^2 - 1)}{T(e^{-2r}\mu +\xi) + 1 - T} &  0 \\
                                                             0 & \mu \\
\end{smallmatrix} \right].\\
\end{align*}

The Holevo quantity can be determined from the symplectic eigenvalues of these two matrices, $\Gamma_{AB}$ and $\Gamma_{A|b}$. If $r=\ln{\sqrt{\mu}}$ then $\chi_{EB}\to0$ in a pure-loss channel.

\subsection{\label{sec:appendix:asymmetricattack}Asymmetric attack}

Now we write down the covariance matrix for general asymmetric channels. After the asymmetric channel, the covariance matrix shared by Alice and Bob is:
\begin{align*}
\Gamma_{AB}^{'} &= \left[ \begin{smallmatrix} \mu & 0 & \sqrt{T_x e^{-2r}(\mu^2-1)} & 0\\ 0 & \mu & 0 & -\sqrt{T_p e^{2r}(\mu^2-1)} \\ \sqrt{T_x e^{-2r}(\mu^2-1)}  & 0 & T_x (e^{-2r}\mu+\xi_x)+1-T_x & 0 \\ 0 & -\sqrt{T_p e^{2r}(\mu^2-1)} & 0 &T_p (e^{2r}\mu+\xi_p)+1-T_p\end{smallmatrix} \right],
\end{align*}
where $r$ is the squeezing parameter in the EB version.  $T_x$ and $\xi_x$ are, respectively, the channel transmittance and excess noise, estimated by Alice and Bob in the $x$ quadrature. However, since the $p$ quadrature is not modulated, $T_p$ and $\xi_p$ are unknown.





\section{\label{sec:appendix:heralding}Covariance matrix heralding protocol}

The covariance matrix shared between Alice and Bob for our heralding protocol can be calculated from the entanglement-based scheme shown in \cref{fig:QKDschemesheralding}. All the elements are Gaussian so it is a simple matter of applying the appropriate symplectic transformations. The covariance matrix after the channel with transmissivity $T$ and excess noise $\xi$ and also after Alice performs her heralding homodyne measurement (assumed to be in the $x$ quadrature) is given by

\begin{align*}
\Gamma_{A1A2B} &= \begin{bmatrix}   
                  a_1 &    0 &              c_1 &    0 &         c_2 &         0 \\
  0 &    {\mu} &       0 &          0 &         0 &            c_3 \\
               c_1 &      0 &   a_2 &     0 &      c_4 &     0 \\
 0 &      0 &    0 &   {\mu} &     0 &   c_5 \\
 c_2 &     0 & c_4 &     0 & b_1 &    0 \\
  0 & c_3 &   0 & c_5 &   0 & b_2 \\
  \end{bmatrix},
\end{align*}
where
\begin{align*}
a_1 &= \frac{e^{4r}{\mu}^2 + 1}{{\mu}(e^{4r} + 1)}\\
a_2 &= \frac{{\mu}^2 + e^{4r}}{{\mu}(e^{4r} + 1)}\\
b_1 &= \frac{e^{4r} - T + T\xi - Te^{4r} + T\xi e^{4r} + 2T{\mu}e^{2r} + 1}{e^{4r} + 1}\\
b_2 &= T\xi - T + \frac{1}{2}T({\mu}e^{-2r} + {\mu}e^{2r}) + 1\\
c_1 &= \frac{e^{2r}({\mu}^2 - 1)}{{\mu}(e^{4r} + 1)} \\
c_2 &= \frac{\sqrt{2T}e^{3r}\sqrt{\mu^2-1}}{e^{4r} + 1}\\
c_3 &= \frac{-\sqrt{2T}e^{r}\sqrt{\mu^2-1}}{2} \\
c_4 &= \frac{\sqrt{2T}e^{r}\sqrt{\mu^2-1}}{e^{4r} + 1}\\
c_5 &= \frac{-\sqrt{2T}e^{-r}\sqrt{\mu^2-1}}{2}.\\
\end{align*}

The Holevo bound can be calculated from the symplectic eigenvalues of the covariance matrix via~\cref{eq:HOLEVO} of the main text, i.e., $\chi_{EB} = S(\rho_{A1A2B}) - S(\rho_{A1A2|B})$.

%


\end{document}